\documentstyle[12pt,aaspp4]{article}
\def\IZw18{I~Zw~18}
\def\m82{M82}

\def\deg{\mbox {$^{\circ}$}}
\def\msun{\mbox {${\rm ~M_\odot}$}}
\def\mstar{\mbox { ${\rm \dot{M}_{*}}$ }}

\def\lsun{\mbox {${~\rm L_\odot}$}}
\def\msunyr{\mbox {$~{\rm M_\odot}$~yr$^{-1}$}}
\def\myk2{\mbox {$~{\rm M_\odot}$~yr$^{-1}$~kpc$^{-2}$}}

\def\Ha{\mbox {H$\alpha$~}}

\def\kpc3{\mbox {${\rm\ kpc}^{3}$}}
\def\o3hb{[OIII]$\lambda5007$~/~H$\beta$~}
\def\O1ha{[OI]$\lambda6300$~/~H$\alpha$~}

\def\s2ha{[SII]$\lambda\lambda6717,31$~/~H$\alpha$~}
\def\2z2{HeII~$\lambda4686$~}
\def\z7{[NII]~$\lambda6583$ }
\def\N2{[NII]~$\lambda6583$~/~H$\alpha$~}
\def\16z2{[SII]~$\lambda\lambda6717, 6731$ }

\def\n{NGC~}
\def\asec{\ifmmode {'' }\else $''~$\fi}  
\def\amin{\ifmmode {' }\else $'~$\fi}    
\def\arcsper{\ifmmode \rlap.{'' }\else $\rlap{.}'' $\fi} 
\def\arcmper{\ifmmode \rlap.{' }\else $\rlap{.}' $\fi} 
\def\sles{\lower2pt\hbox{$\buildrel {\scriptstyle <}
   \over {\scriptstyle\sim}$}} 
\def\sgreat{\lower2pt\hbox{$\buildrel {\scriptstyle >}
    \over {\scriptstyle\sim}$}} 
\def\kms{~km~s$^{-1}$~}

\def\cm3{~cm$^{-3}$}

\def\fig{{Figure}}

\def\x{{X-ray}~}

\def\et{{\rm et\thinspace al.}\ }   

\def\apj{ApJ}
\def\apjs{ApJS}

\def\aj{AJ}
\def\mn{MNRAS}

\def\aa{A\&A}

\begin{document}

\title{Properties of Galactic Outflows:  Measurements \\
of the Feedback from Star Formation}

\author{Crystal L. Martin\altaffilmark{1,2}}

\altaffiltext{1}{Hubble Fellow}

\altaffiltext{2}{Space Telescope Science Institute, 3700 San Martin
Drive, Baltimore, MD 21218}


\begin{abstract}
Properties of starburst-driven outflows in dwarf galaxies are
compared to those in more massive galaxies. Over a factor of $\sim 10$ in 
galactic rotation speed, supershells are shown to lift warm 
ionized gas out of the 
disk at rates up to several times the star formation rate. 
The amount of mass escaping the galactic potential, in contrast to
the disk, does depend on the galactic mass.  The temperature of 
the hottest extended \x emission shows little variation around 
$\sim 10^{6.7}$~K, and this gas has enough energy to escape from the galaxies 
with rotation speed less than approximately $130^{+20}_{-40}$\kms.
\end{abstract}


\section{Introduction}
Most models of galaxy formation and evolution contain
a critical parameter called {\em feedback}.  It describes the
efficiency with which massive stars reheat the surrounding interstellar
medium (ISM) and is thought to have a particularly strong impact on the
star formation history of low mass galaxies (Dekel \& Silk 1986, Larson 1974).
In CDM-based models for the hierarchical assembly of galaxies, 
strong {\em differential feedback} seems to be required to
reproduce the observed galaxy luminosity function and the mass --
metallicity relation among galaxies
(Kauffman, Guiderdoni, White 1994;
Cole \et 1994; Somerville \& Primack 1998). 
Given the growing recognition of the importance of feedback and the 
spatial resolution limits of numerical simulations, empirical
descriptions on scales $\sgreat\ 1$~kpc are needed. Relevant observations of 
the warm and hot ISM in nearby galaxies are compiled here,
and the implications for feedback recipes and galaxy evolution are discussed.





\section{Data:  Galaxies with Strong Feedback}
\label{sec:sample}



A sample of dwarf, spiral, and starburst galaxies was constructed from the
literature on galactic winds and extraplanar, diffuse ionized
gas (DIG).  Most of these galaxies have at least one region where the surface
brightness approaches the empirical limit of 
$L \approx 2.0 \times 10^{11}\lsun $~kpc$^{-2}$ 
(Meurer \et 1997).  Assuming the slope 
of  the stellar initial mass function (IMF) is Salpeter ($\alpha = 2.35$), the
corresponding star formation rate of 1 to 100\msun\ stars
is  $\mstar \sim 14$\myk2. This scale, Leitherer \& Heckman (1995), is used for
 the star formation rate (SFR) throughout this paper, and  extending the 
IMF to 0.1\msun would increase the SFRs by a factor of 2.55.

\subsection{Dwarf Galaxies}

Large expanding shells of warm ionized gas are common in dwarf galaxies
with starburst, i.e. high surface brightness,  regions
(e.g. Hunter \& Gallagher 1990; 
	Meurer \et 1992; 
	Marlowe \et 1995; 
	Hunter \& Gallagher 1997; 
	Martin 1998), 
but these galaxies are not particularly representative of the dwarf galaxy 
population. The local number density of dwarf galaxies is 
sensitive to a survey's surface brightness limit (e.g. Dalcanton 1997)
and unknown at the level of a factor of at least 2 to 3.
Samples that include some lower surface brightness dwarf irregular
galaxies,  Hunter, Hawley, \& Gallagher (1993, HHG) for example, have
mean $M_{HI} / L_{H\alpha}$ about 1 dex higher than
samples of blue amorphous dwarfs (Marlowe \et 1995).
While extraplanar, expanding filaments were found in 7 of 12 galaxies in
the latter sample, only 2 of the 15 
galaxies  with inclinations $i > 60\deg$ in the HHG sample even 
contain {\em extended} filaments.  Even the HHG sample might be 
missing more than 1/2 the dwarfs, so 
the fraction of nearby dwarf galaxies currently in an outflow stage is
unlikely to be more than 5\%.  Only galaxies with high star formation rates 
per unit area are discussed in this paper.  
The frequency of expanding shells is similar to the  Marlowe \et sample,  
but a broader range of morphological types is included.
The absolute magnitude
of the galaxies ranges from  $M_B \approx -13$ to $M_B = -18.5$ and
$0.84 < (M_{HI} / L_{H\alpha}) / (\msun/\lsun) < 3.17$. 
Expanding shells  were detected in 12 of 14 
galaxies using \Ha longslit,  echelle spectra, and the filaments
are clearly extraplanar in 6 galaxies (Martin 1998).


Star formation rates for these dwarf galaxies were derived from the integrated 
\Ha fluxes after correcting for Galactic extinction (Paper I).
The intensity of large star forming complexes in many of the dwarfs with
strong \Ha emission and extraplanar emission  reaches several \myk2.
 Averaged over the optical area of a galaxy 
(i.e. $\pi R_{25}^2$), however, a typical star formation rate 
is $1.14 \pm 0.11 \times 10^{-3}$\myk2. 
Only two galaxies, \n1569 and \n4449, have secure detections of 
extended, thermal X-ray emission (Della Ceca \et 1996, 1997).  X-ray emission
has been detected from several others -- \n5253, \n4214, \n1705, IZw18, and VIIZw403, but the thermal emission is not unambiguously resolved from point sources.
The peculiar galaxy M82, which is not much more luminous than a dwarf galaxy,
also has an \x halo (e.g. Strickland \et 1997).



\subsection{Comparison Sample}

Spiral disks with high areal star formation rates show extraplanar 
DIG (cf. Table~2 in Rand 1996). The typical spiral galaxy has DIG in the 
spiral arms, but extraplanar plumes are only present above particularly 
active sites of localized star formation (Walterbos \& Braun 1994;  Wang \& 
Heckman 1997).  This paper examines the DIG in 6 edge-on galaxies
with $L_{FIR} / D_{25}^2 > 1 \times 10^{40}$~ergs/s/kpc$^2$,  or star
formation rates greater than $2 \times 10^{-4}$\myk2.
The far-infrared luminosity  was adopted as the star formation indicator
since extinction corrections dominate the \Ha luminosity for these galaxies,
but the uncertainties in the SFR may be as large as a factor of two
(e.g. Sauvage \& Thuan 1992). 

%


Measurements of \x halo properties are drawn from Dahlem, Weaver, and 
Heckman (1998, DWH) who have combined the available ROSAT and ASCA data for a 
flux limited sample of nearby edge-on starburst galaxies.
It is not surprising that some of these galaxies are common to Rand's sample, 
since filaments protruding from the nucleus account for much of the extended 
DIG emission.  The mean $L_{IR} / D_{25}^2$ of the starburst sample, 
$\sim 2.6 \pm 4.0 \times 10^{-3}$\myk2, is about an order of magnitude higher 
than that of Rand's sample.  More meaningful concentration indices like
$L_{IR} / \pi R_{e,H\alpha}^2$ give areal star formation rates of order one 
\myk2\ but are often not well-defined due to severe extinction 
(Lehnert \& Heckman 1996). 
Much (25\% to 78\%) of the massive star formation in the local universe 
takes place within the central $\sim 1~$kpc of galaxies like these
(Gallego \et 1995; Heckman 1998), and 
the emission line widths, shock-like line ratios, and \Ha morphology 
demonstrate that minor-axis outflows are prevalent (Lehnert \& Heckman 1995).



%
%

\section{Results}
\label{sec:results}

\subsection{Disk Mass Loss Rates in Dwarf Galaxies}
\label{sec:mwim}

%

Prominent shells and filaments are plainly visible in the \Ha imagery
of the star-forming dwarf galaxy sample.
The subset of filaments which comprise large, expanding shells were
kinematically identified in Paper~I, and their \Ha luminosities are
tabulated in Table~3 of Paper~I.  The density in the extended filaments
$n_e$, is too low to measure with common line ratio diagnostics (e.g. 
Osterbrock 1989), so shell masses were parameterized in terms of an 
unknown volume filling factor $\epsilon$, where $n_{rms}^2 = \epsilon n_e^2$.
For any measured luminosity, condensations along the sightline reduce the 
inferred mass, $M \propto \epsilon^{1/2}$, but increase the inferred
pressure,  $P \propto \epsilon^{-1/2}$.  As illustrated in 
Figure~\ref{fig:MP}, varying the volume filling factor, $\epsilon$, from 1 
(upper left) to $10^{-5}$ (lower right) changes the inferred mass of the 
largest shells from several times $10^6$\msun to $10^4$\msun.  
The pressure in the warm filaments is unlikely, however,  to exceed the 
pressure of the hot gas which presumably fills the interior cavity.  

For one of the nearest starburst galaxies, \n1569, extended, soft \x emission  
was identified
in ROSAT hardness maps,  and a soft, thermal component was required to
fit the integrated ASCA spectrum (Della Ceca \et 1996).  
The thermal pressure derived from this model depends more on 
the choice of plasma model and shell volume than any calibration uncertainties,
and the acceptable range is illustrated in \fig~\ref{fig:MP} by vertical lines.
The Meka thermal model with the large, H95 volume gives the lowest $P_x$,
while the \Ha shell volume plus Raymond-Smith spectral fit allows a pressure
roughly 4 times higher. 
Pressure equilibrium between the hot and warm gas
requires $ -3 < \log \epsilon < -2$ for each of the large shells protruding 
from \n1569.


%
%
%

The same argument implies $\log \epsilon \approx -2$
for the large \Ha shells associated with the extended \x emission from \n4449.
\fig~\ref{fig:MP}b illustrates the substantial range of pressure allowed by 
the volume estimates. The pressure of the very soft thermal component 
(0.24 keV) is similar to, actually 1/2 as large as, that of the soft component 
(0.82 keV).  For M~82, \fig~\ref{fig:MP}c, filling factors from $10^{-4}$ to 
$10^{-2}$ would bring the pressure of the warm ionized filaments into 
the $P_x$ range measured along the outflow (Strickland \et 1997). 
The filling factor seems to be about 10 times lower in the M~82 
outflow than in \n1569 and \n4449, and a similar difference has been
measured for their HII region filling factor (Martin 1997, Table 6).
In contrast, the very high filling factor inferred for \n4449 
from the DGH97 volume estimate would be difficult to reconcile with
the very low HII region ionization parameter.

The gas pressures in \fig~\ref{fig:MP} are high compared to the local Milky 
Way ISM but quite reasonable.  
Adiabatic bubble models for the shells' expansion predict pressures
of $2.0 \pm 1.4 \times 10^5~k$~K \cm3, $1.3 \pm 2.2 \times 10^5~k$~K\cm3, and
$7.2 \pm 1.4 \times 10^5~k$~K\cm3 in \n1569, \n4449, and M~82 respectively.
(These values assume a mean ambient density of $n_0 \approx 0.1$\cm3, use
the power and ages from Table~2 of Paper~I, and assign 5/11 of the energy
to the hot, shocked gas (Koo \& McKee 1992).) 
The magnetic pressure, $ B^2/8\pi $, is 1 to 2 orders of 
magnitude smaller than this  in \n1569, $4.2 \times 10^4~k$~K~cm$^{-3}$ 
(Israel \& deBruyn 1988), 
and in M~82, $ 3.5 \times 10^3~k$~K~cm$^{-3}$ (Seaquist 
\& Odegard 1991). The filling factors derived from the pressure equilibrium
argument are therefore expected to be accurate to better than a factor of 10.
Based on these results, 
the warm ionized gas masses in Tables~3 and~5 of Paper~I would be more
revealing if parameterized in terms of $\epsilon = 0.01$ rather than the
original $\epsilon = 0.1$. 
The corrections to the mass obtained using the $\epsilon $ values derived 
above for \n1569, \n4449, and M~82 are then minor.  For a particular galaxy,
a lower limit on the disk mass loss rate is simply the sum of its
shell masses divided by the age of the oldest shell,

\subsection{Reheating Efficiency}
\label{sec:eff}

\fig~\ref{fig:main} shows the ratio of disk mass loss rate, $\dot{M_w}$,
to star formation rate as a function of circular velocity.  
Although the warm ionized shells typically contain at most a few percent of 
the galactic gas mass, they lift gas out of the disk at rates comparable to
the rate gas goes into new stars. No trend is seen with $V_c$ over the 
luminosity/mass interval of the dwarf sample -- the solid symbols.

Comparison of the reheating efficiency measured in the dwarf galaxies
to that in more massive disk galaxies is not straightforward.  
The mass of the extended DIG in the spirals \n4013, \n4302  and \n3079 
was computed from emission profiles -- 
$n_{rms}^2(R,z) = <n_{rms}^2>_0 e^{-z/z_0}$ for $R \le R_0$ -- fit to deep
\Ha images (Rand 1996; Veilleux \et 1995). The halo DIG mass for \n891 is from Dettmar (1990). For \n4631,
the hot gas mass loss rate, $\dot{M}_x$, from Wang \et (1995) 
was substituted for $\dot{M}_{wim}$. 
All measurements were scaled to a common filling factor
of $\epsilon = 10^{-2}$ for comparison to the dwarf galaxies sample, but 
measurements of the latter were not
corrected for [NII] emission in the filter bandpass.  
The gas dynamical timescale was set equal to
the emission measure scale height divided by the sound speed at $10^4$~K. 
 Open symbols in 
Figure~\ref{fig:main} show the resulting ratio,  $\dot{M}_{w} / \mstar$, 
for these five galaxies.
Any point in this diagram is uncertain by a factor of 2-3, but it is 
remarkable that  -- over  a factor of nearly 10 in galactic rotation speed --
the upper envelope shows little variation around $\dot{M} / \dot{M}_{*} 
\sim 5$. This upper limit probably indicates something fundamental 
about the reheating efficiency.  In particular, it is  more related to the 
areal density of stars than the depth of the potential.



\subsection{Galactic Mass Loss}
\label{sec:mx}

%
%

The fate of the gas in the expanding shells depends on the gravitational 
potential of the galaxy.  For a measured rotation speed (Paper~I),
the distribution of matter in both the galactic disk  and the halo affect the
estimated depth. For example,
the escape velocity at  $R(max~ V_c)$ is at least $1.414 V_c$ but 
increases to $3.55 V_c$ or 2.57 $V_c$ for spherical, isothermal halos
extending, respectively, to 100 times or 10 times this radius.
The shells in \n1569, one shell in Sextans~A, and one shell \n3077
have projected expansion speeds greater than $1.414 V_c$; but
only one of the shells in \n1569 is expanding faster than $3.55 V_c$.
Hence, even in dwarf galaxies, much of the warm, ionized gas blown out of a disk 
probably remains bound to the galaxy.

The fate of the hot gas confined by the shells may be different.
 Supershells accelerate when they reach several gas scale heights and
break up from Rayleigh-Taylor instabilities (MacLow, McCray, \& Norman 1989).
The hot, interior gas exits at the sound speed.  In the absence of radiative 
losses, gas hotter than $T_{esc} = 1.5 \times 10^5 (v_{esc}/100 {~\rm km/s})^2$
escapes the galactic potential.  This critical temperature represents
a specific enthalpy equal to $1/2 v_{esc}^2$.  \fig~\ref{fig:tv} shows its
variation with galactic rotation speed for the three $v_{esc} / V_c$ 
ratios discussed above.  The temperature of the hot gas in \n1569 and \n4449 
is well above all these limits. The temperature of the M~82 outflow also
exceeds the escape temperature if the halo is severely truncated  --
i.e. the bold line (see Sofue \et 1992). 
Solar metallicity gas at $T = 10^{6.8}$~K and
$n = 0.01$\cm3 cools radiatively in $\sim 2 \times 10^8$~yr (Sutherland \& 
Dopita 1993), and the halo gas could reach a radius of $\sim 40$~kpc in
this time.
The mass of the hot outflow in \n1569 is $M_x = 6.12 - 6.99 \times 10^5 \msun\ 
\sqrt{V / 1 {\rm\ kpc}^3} $, or $6.3 - 7.2 \times 10^5 \msun$; and the soft and 
very soft components in \n4449 contain $M_x = 5.3 \times 10^5 \msun$ and 
$M_x = 7.9 \times 10^5 \sqrt{V / 1 {\rm\ kpc}^3} \approx 8.9 \times 10^5 \msun$ 
respectively.  The X-ray emitting gas contains about as much mass as the 
\Ha shells, so 
the disk mass loss rates in \fig~\ref{fig:main} are indicative of 
the galactic mass loss rate as well.

The importance of this result for modeling feedback is amplified by 
measurements of $T_x$ in more massive galaxies.  The temperature constraints
found for \n891 (Bregman \& Houck 1997) and \n4631 (Wang \et 1995) are shown
in \fig~\ref{fig:tv} along with the sample re-analyzed by Dahlem \et (1998).
The temperature of the hot gas in these galaxies is similar to that in the 
two dwarf galaxies and M82, about $T_x \sim 10^{6.8}$~K.  Although
foreground Galactic absorption could hide a lower temperature thermal component in several of these galaxies, the ASCA spectra which extend to 10~keV would 
have detected a hotter thermal component if it were present.  Since the minimum
in the cooling curve occurs at a higher temperature, $T \approx 10^{7.4}$~K, 
the temperature uniformity must reflect the reheating efficiency of massive 
stars. As illustrated in \fig~\ref{fig:tv}, the escape temperature from an 
extended halo 
rises above the hot gas temperature at a circular velocity $\sim 130$\kms.
The hot gas in the outflow is therefore expected to form a bound halo around
larger galaxies.


\section{Discussion: Recipes and Implications}
\label{sec:discuss}
%

To describe the global impact of star formation on the ISM, a very simple 
empirical feedback recipe is proposed. Three components of interstellar gas, 
which can be referred to as cold, warm, and hot must be identified.
Use the Schmidt law parameterization of Kennicutt (1998) to estimate
the global SFR.  If the SFR averaged over the area of
the stellar disk exceeds a few times $10^{-4}$\myk2, then transfer warm 
($\sim 10^4$~K) disk gas to the halo at a rate of a few times the star formation rate.  
Generate hot, $\sim 10^{6.7}$~K, gas at a similar rate and remove it from the halo
if the rotation speed is less than $\sim 130^{+20}_{-40}$\kms.


Comparison of this empirical recipe to those in the semi-analytic galaxy
formation models (SAMs) of the Munich, Durham, and Santa Cruz  groups 
provides some insight into the impact of such a recipe. The observations 
indicate that the {\em differential} aspect of the feedback is
the escape fraction of hot gas from the halo. For simplicity,
the empirical recipe presents a sharp transition from ejection to retention,
but a milder increase in ejection efficiency toward lower circular velocity
could still be quite reasonable.  The disk reheating rate was found to
be insensitive to $V_c$ in contrast to the common prescription in the
SAMs where $\dot{M}_{reheat} / \mstar \propto V_c^{\alpha}$ with $\alpha 
\sim -2$.  All three groups enhance the reheating efficiency in
the dwarfs, but the temperature assigned to this reheated gas, or 
equivalently its fate, differ.  If the reheated gas is ejected
from the halo (e.g. Cole \et 1994; SP98), the prescription  becomes very
similar to the empirical recipe.  These ejection models flatten
the faint-end of the luminosity function more than models which retain
the reheated gas (e.g. KGW94).   
The empirical feedback recipe will not, however, suppress star formation
in small halos as strongly as the Durham prescription.  The latter
lowers the star formation effeciency in dwarfs in addition to increasing
the feedback,  and Figure~7 of SP98 indicates this causes too much 
curvature at the faint end of  the Tully-Fisher relation. 
The mass -- metallicity relation depends on 
assumptions about the composition of outflow but would seem to be in a
reasonable regime (e.g. Figure~13 of SP98).


The outflows observed in nearby dwarf galaxies do not expel the entire disk
over the lifetime of individual starburst regions.  For example, 
the wind in \n1569 might expel $\sim 0.3$\msunyr\ over $10^8$~yr, or $3 \times
10^7$\msun\ of the disk.  If this mass is swept out of the central cylinder
of radius 500~pc and height 1~kpc, the concentration of the ejected disk 
material is $M_d / r_d = 0.07 V_{30}^{-2}$ in units of the halo mass to 
scalelength $M_h/a_h$. Much larger concentrations, like $M_d/r_d \sim 1 - 20$,
must be ejected to unbind a substantial amount of the central cusp in 
the dark matter distribution (Navarro, Eke, \& Frenk 1996).  The mass
lost from local star-forming dwarfs does not seem to be sufficient to 
generate the dark matter cores observed in some low surface brightness 
dwarf galaxies. If areal SFRs were $\sgreat\ 10$ times higher in halos of 
a given circular velocity at high redshift, then the fraction of warm and 
cold gas escaping the halo would have been more significant. This fraction 
would also be increased if environmental effects both trigger starbursts 
and truncate their surrounding dark matter halos (e.g. M82, Sofue \et 1992).  
It is unclear, however, whether the reheating could have smoothed out the
gas distribution enough to prevent the severe angular momentum losses that
plague current N-body/gasdynamical simulations of galaxy formation 
(Navarro \&  Steinmetz 1997).

%
%
%

%

The empirical recipe can be improved with further work. Only the most
vigorously star-forming local galaxies were considered in this paper,
but the critical areal SFR for supershell blowout and its sensitivity to
HI scale height could be measured.  The total mass at temperatures, $\sim 10^5
 - 10^6$~K, needs to be better constrained. 	
It is  similar to that in the hot $10^{6.7}$~K phase 
for  \n4449 and \n4631 -- two galaxies with foreground absorption
low enough to allow detection in the \x spectrum.
Both molecular and neutral atomic gas have been detected in galactic outflows
(Sofue \et 1992; Toshihiro \et 1992; Heckman \& Leitherer 1997),
but the ubiquity of a cold component and its mass need to be determined.
The biggest systematic uncertainty affecting the galactic mass loss rates
is radiative losses. Mass loaded outflows could
radiate more of the thermal energy reservoir than assumed here, so
a better understanding of the transfer of mass and energy between different 
phases of gas in the outflows is needed.
Feedback can be countered to some degree by
adjusting cosmological parameters, particularly the 
slope of the power spectrum on small scales (SP98).  
Tighter empirical constraints on the feedback would help  ensure SAMs 
narrive at the physical solution.

\acknowledgements{
Support for this work was provided by NASA through Hubble Fellowship grant
\#HF-01083.01-96A awarded by the Space Telescope Science Institute, which is 
operated by the Association of Universities for Research in Astronomy, Inc., 
for NASA under contract NAS 5-26555.  This work benefited from discussions
with Carlos Frenk, Tim Heckman, Guinevere Kaufmann, Cedric Lacey, 
Julio Navarro, Rachel Somerville,  and Matthias Steinmetz 
at the Aspen Center for physics.  Literature searches were performed with NED,
the NASA/IPAC Extragalactic Database, a facility operated by the Jet
Propulsion Laboratory, Caltech, under contract with NASA.
}

%
%

\newpage


	\begin{figure}[h]
	\caption{}
	Mass and pressure of the warm, ionized shells in 3 actively
	star-forming galaxies.
  	The diagonal lines illustrate the effect of varying the volume filling
	factor $\epsilon$ from unity (upper left) to $10^{-5}$
	(lower right) in increments of 1 dex.  Vertical lines denote
	the pressure of the hot bubbles.  Allowed range given by:
		(a) Mewe and Raymond-Smith models (DGHM96) with
	            volumes from 1.061~\kpc3 (Paper~I) to 4.188 \kpc3\ 
		    (Heckman \et 1995),
		(b) Meka and Raymond-Smith models (DGH97) with
		    volumes from 1.290\kpc3 (Paper I) to 9.4\kpc3 
		    (DGH97 scaled to d = 3.6~Mpc).
		(c) Gradient along outflow (Strickland \et 1997).
	\label{fig:MP}
	\end{figure}

	\begin{figure}[h]
	\caption{
	Mass ejection efficiency versus maximum HI rotation speed.  
	Filled circles represent the ratio of the WIM mass loss rate,
	which scales as $\sqrt{\epsilon / 0.01}$, to the galactic SFR
	in a sample of high SFR dwarf galaxies.  
	Arrows illustrate corrections to the filling factor derived from
	pressure equilibrium arguments.  Open symbols show an analogous
	ratio for spiral galaxy disks with high areal star formation rates.
	Pairs of triangles denote upper and lower limits for a galaxy.
	}
	\label{fig:main}
	\end{figure}

	\begin{figure}[h]
	\caption{}
	Temperature of extended, thermal X-ray emission versus maximum HI
	rotation speed.  Solid symbols denote a second thermal component
	when detected. Shown are \n1569 (Della Ceca \et 1996, $\Box$), 
		\n4449 (Della Ceca \et 1997, $\Box$), 
		M~82 (Strickland \et 1997, $\triangle$; DHW, $\bigcirc$),
		\n4631 (Wang \et 1995, $\triangle$)
		\n891 (Bregman \& Houck 1997, $\triangle$),
		\n253, \n3079, \n3628, and \n2146 (DWH, $\bigcirc$).
        The solid line illustrates the minimum escape temperature -- i.e. 
	all the mass is interior to the location where the rotation speed 
	was measured.  The dotted lines show the escape temperature from 
	isothermal halos truncated at 10 times and 100 times this radius.
	\label{fig:tv}
	\end{figure}
\vfill\eject\

%

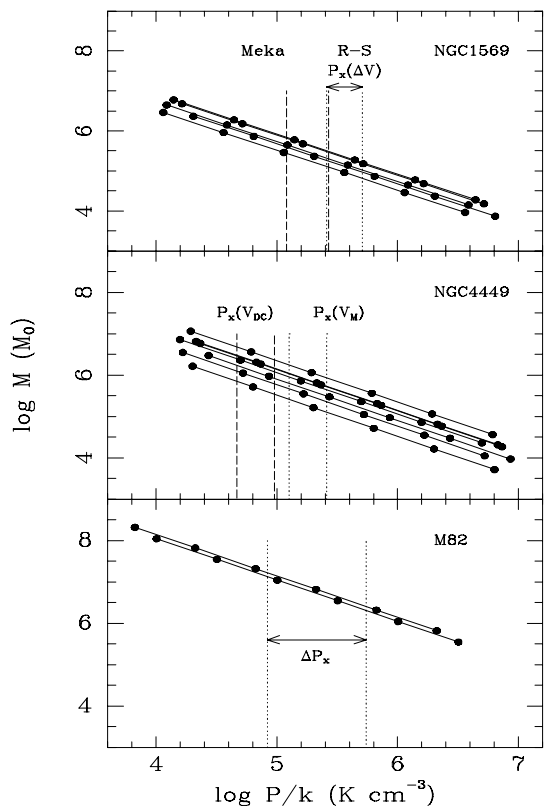
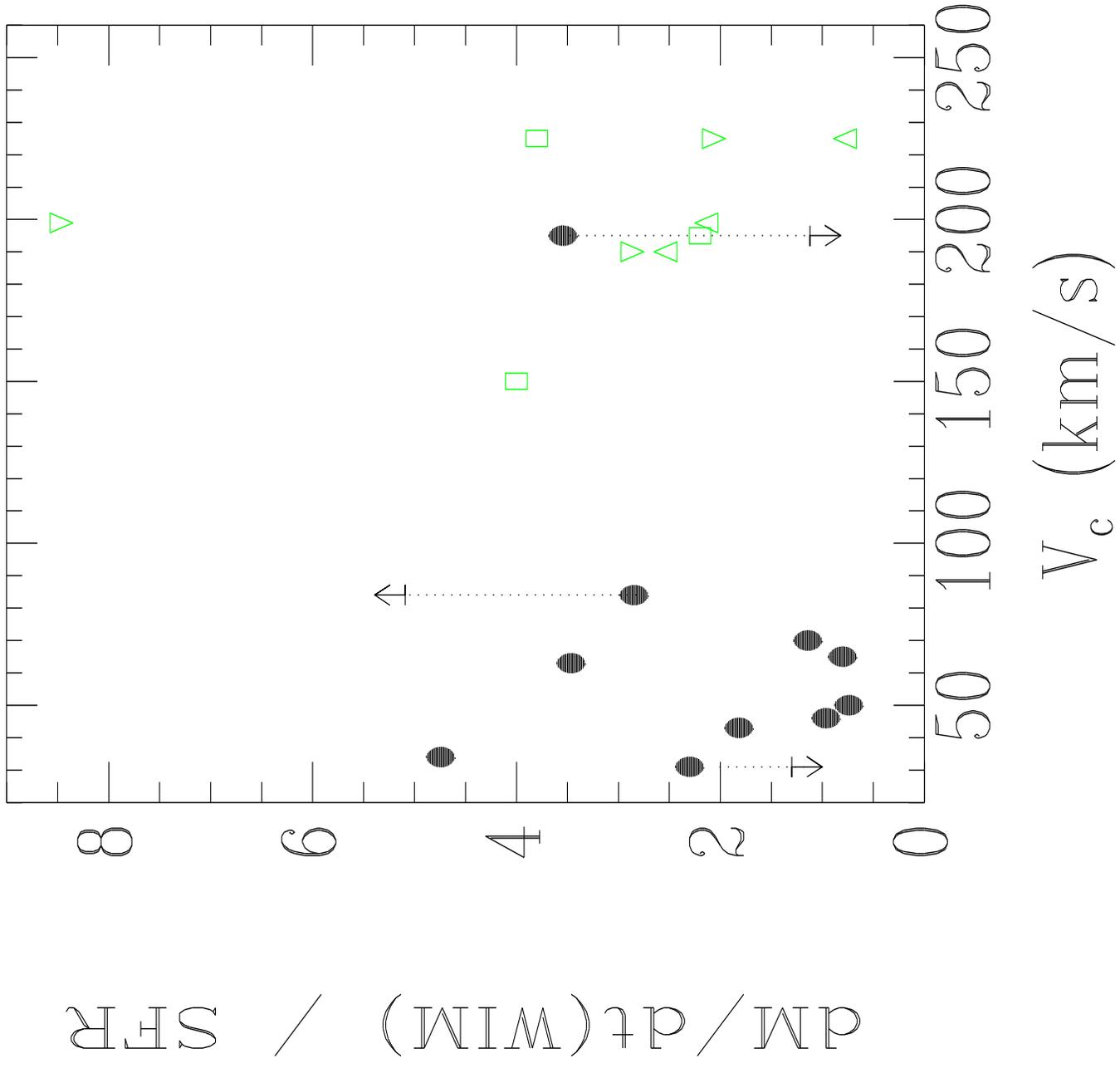
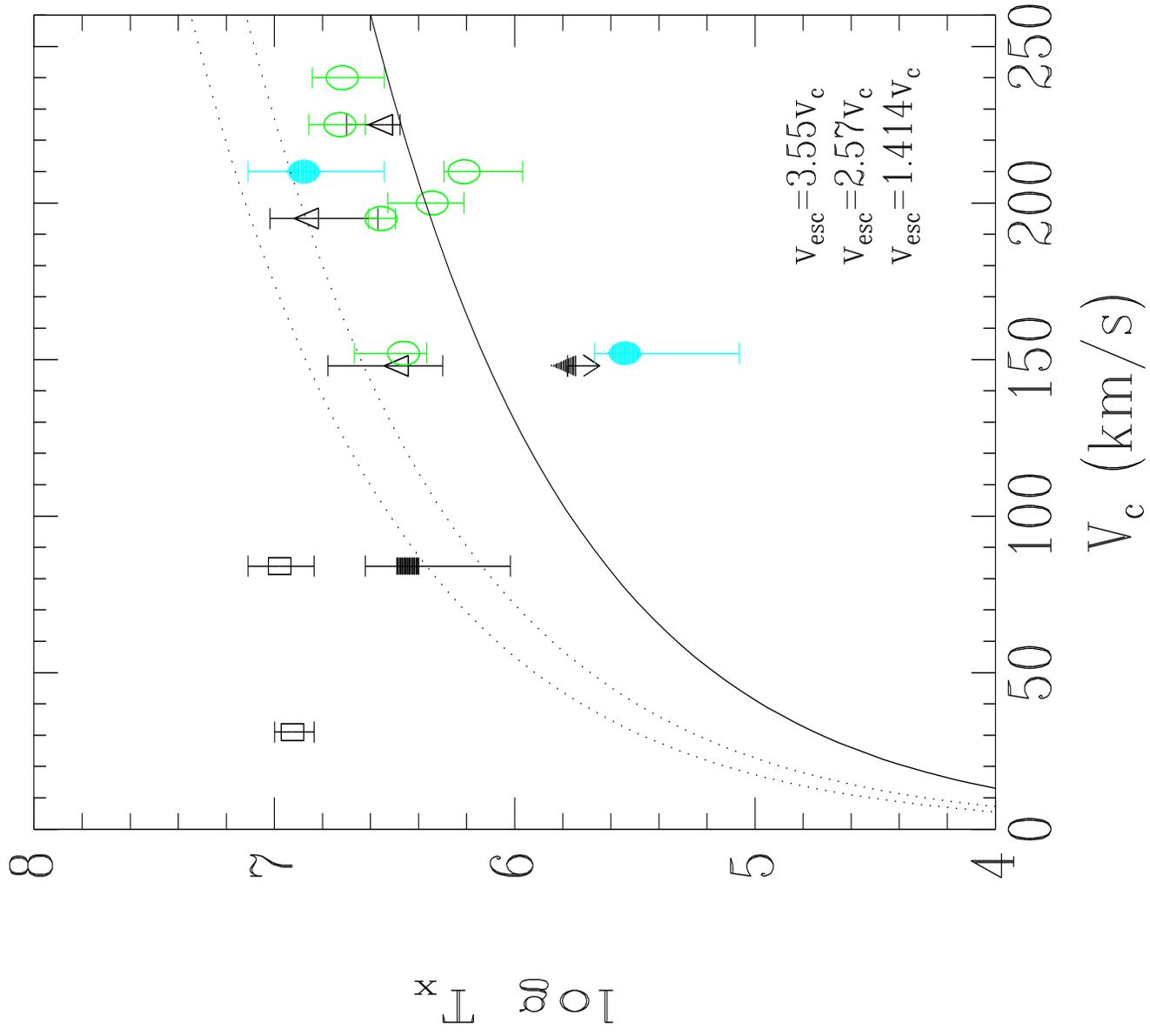
\end{document}